\documentclass[runningheads]{llncs}

\usepackage{graphicx}
\usepackage{hyperref}
\usepackage{booktabs}

\usepackage{wrapfig}
\usepackage{siunitx} % make table numbers layout nicer
\newcommand{\datasetname}{SAP Signavio Academic Models}
\newcommand{\datasetacr}{SAP-SAM}
\newcommand{\saiacr}{SAP-SAI}
\newcommand{\mypar}[1]{\noindent\textbf{#1.}}

\usepackage{xcolor}

\newcommand{\mycomment}[1]{}
\begin{document}
\title{\datasetname: A Large Process Model Dataset}
\titlerunning{\datasetacr: A Large Process Model Dataset}
\author{Diana Sola\inst{1,2}\orcidID{0000-0001-5688-1730} \and
Christian Warmuth\inst{1}\orcidID{0000-0003-0125-1824}
\and Bernhard Schäfer\inst{1,2}\orcidID{0000-0003-4364-0086}
\and Peyman Badakhshan\inst{1}\orcidID{0000-0002-7627-7618}
\and Jana-Rebecca Rehse\inst{2}\orcidID{0000-0001-5707-6944}
\and Timotheus Kampik\inst{1}\orcidID{0000-0002-6458-2252}
}
\authorrunning{D. Sola et al.}
\institute{SAP Signavio, Berlin, Germany \and University of Mannheim, Mannheim, Germany \\  % Data and Web Science Group, TK: can we simplify affiliations?
\email{\{diana.sola,christian.warmuth,bernhard.schaefer,\\peyman.badakhshan,timotheus.kampik\}@sap.com, rehse@uni-mannheim.de}}
\maketitle              % typeset the header of the contribution
\begin{abstract}
In this paper, we introduce the \emph{\datasetname} (\datasetacr) dataset, a collection of hundreds of thousands of business models, mainly process models in BPMN notation. The model collection is a subset of the models that were created over the course of roughly a decade on \url{academic.signavio.com}, a free-of-charge software-as-a-service platform that researchers, teachers, and students can use to create business (process) models.
We provide a preliminary analysis of the model collection, as well as recommendations on how to work with it.
In addition, we discuss potential use cases and limitations of the model collection from academic and industry perspectives.

\keywords{Process Models  \and Data Set \and Model Collection.}
\end{abstract}
%
%
%
%%%%%%%%%%%%%%%%%%%%%%
\section{Introduction}
%%%%%%%%%%%%%%%%%%%%%%
Process models depict how organizations conduct their operations. They represent the basis for understanding, analyzing, redesigning, and automating processes along the business process management (BPM) lifecycle~\cite{DumasLaRosaEtAl13}. 
As such, many organizations posses large repositories of process models~\cite{houy2011business}. 
Having access to such repositories would be tremendously beneficial for developing and testing algorithms in the area of BPM, e.g., for process model querying \cite{Polyvyanyy2022} or reference model mining \cite{rehse2017graph}.
Also, the growing interest in applying machine learning in the BPM field, e.g., for process model matching~\cite{antunes_2015_matching}, process model abstraction~\cite{wang2018business} or process modeling assistance~\cite{sola2021use}, underlines the relevance for large model collections that can, for example, serve as training datasets.

%However, only a small number of real process models from practice are publicly available. 
However, researchers rarely have access to large collections of models from practice. 
%\tnote{We are not addressing this problem, though :-D Can we change our line or argumentation here?} \dnote{hm... any ideas @Jana @Timmi \& others?} \jnote{I don't think we need to change much. I tweaked it a little, check it out.}
Such models can contain sensitive information about the organization's internal operations.
Legal aspects and the fear of losing competitive advantage thus discourage companies from publishing their business (process) models \cite{thaler2014need}. 
This inherent dilemma has so far largely prevented the publication of large-scale model collections for research, as they are common in related research fields \cite{thaler2014need}.

%The so far available process model datasets are also rather small. For example, hdBPMN~\cite{schaeferSketch2BPMN2021} contains 700 handwritten BPMN 2.0 models that can be parsed as XML files.
%RePROSitory (Repository of open PROcess models and logS) is an open collection of business process models and logs, meaning users can contribute to the repository by uploading additional models and logs. At the time of writing, RePROSitory comprises around 700 models.
% The so far available process model datasets are rather small. With 29,810 models, the model collection of the business process management academic initiative is the largest among them \cite{weske_mathias_2020_3758705}. 

In this paper, we introduce \emph{\datasetname} (\datasetacr), a model collection that consists of hundreds of thousands of process and business models in different notations. 
We provide a basic overview of datasets related to \datasetacr, as well as the origin and structure of it. Subsequently, we present selected properties and use cases of \datasetacr.
Finally, we discuss limitations of the dataset along with recommendations on how to work with it.

%%%%%%%%%%%%%%%%%%%%%%%%%%%%%%%%
\section{Related Datasets}
%%%%%%%%%%%%%%%%%%%%%%%%%%%%%%%%

Compared to \datasetacr, existing process model collections are rather small. The hdBPMN~\cite{schaeferSketch2BPMN2021} dataset, for example, contains 704 BPMN 2.0 models. This collection has the special feature that the models are handwritten and can be parsed as BPMN 2.0 XML. 
Another example is RePROSitory \cite{reprository2019} (Repository of open PROcess models and logS) which is an open collection of business process models and logs, meaning users can contribute to the repository by uploading their own data. At the time of writing, RePROSitory also contains around 700 models.
Some models included in \datasetacr{} have already been published \cite{weske_mathias_2020_3758705}. However, the previously published dataset contains only 29,810 models that were collected over a shorter period of time.

In the process mining community, the BPI challenge datasets, e.g., the BPI challenge 2020 \cite{bpiChallenge2020}, have become important benchmarks. Unlike \datasetacr, these datasets consist of \textit{event logs} from practice. Therefore, the applications of the BPI challenge datasets only partially overlap with those of \datasetacr.

%%% COMMENTS, to be removed
%was extracted from the database
%\begin{itemize}
%    \item write about SAP Signavio academic + short intro of how the data has been collected
%    \item short description of dataset
%    \item write about challenges of collecting and investigating such a big process model dataset and how we tackled them
%    \item paper overview
%\end{itemize}

%%%%%%%%%%%%%%%%%%%%%%%%%%%%%%%%%%%
\section{Origins \& Structure of \datasetacr}
%%%%%%%%%%%%%%%%%%%%%%%%%%%%%%%%%%%

\datasetacr\ contains 1,021,471 process and business models that were created using the software-as-a-service platform of the SAP Signavio Academic Initiative\footnote{See: \url{signavio.com/bpm-academic-initiative/} (accessed at 2022-07-25)} (\saiacr), roughly from 2011 to 2021\footnote{The total number includes vendor-provided example models, which are automatically added to newly created workspaces (process repositories that users register). About 470,000 models in the dataset bear the name of an example model, but this can only be a rough estimate of the number of example models in the dataset.}. 
Most models are in Business Process Model and Notation (BPMN 2.0\footnote{Technically, the latest version of BPMN is, at the time of writing, BPMN 2.0.2. However, little has changed between 2.0 and 2.0.2. We assume that the informal cross-vendor alignment efforts of the OMG BPMN Model Interchange Working Group are more substantial than formal progress between minor versions. In the following, we therefore use \textit{BPMN 2.0} to refer to any version among 2.0 and 2.0.2.}). 
\saiacr\ allows academic researchers, teachers, and students to create, execute, and analyze process models, as well as related business models, e.g., of business decisions.  
The usage of \saiacr\ is restricted to non-commercial research and education. Upon registration, users consent that the models they create can be made available for research purposes, either \emph{anonymized} or \emph{non-anonymized}.
\datasetacr\ contains those models for which users have consented to non-anonymized sharing.
Still, anonymization scripts were run to post-process the models, in particular to remove email addresses, student registration numbers, and---to the extent possible---names.

The models in \datasetacr\ were created between July~2011 and (incl.) September~2021 by a total of 72,996 users, based on a count of distinct user IDs that are associated with the creation or revision of a model.
The models were extracted from the MySQL database of \saiacr\ and are in SAP Signavio's proprietary JSON-based data format.
The BPMN models are conceptually BPMN-2.0-standard-compliant, i.e., individual models can be converted to BPMN 2.0 XML using the built-in functionality of \saiacr.
Decision Model and Notation (DMN) models can be exported analogously.
The dataset contains models in the following notations:

\begin{itemize}
    \item Business Process Model and Notation (BPMN): BPMN is a standardized notation for modeling business processes~\cite{omg2013bpmn202}. \datasetacr\ distinguishes between BPMN process models, collaboration models, and choreography models, and among BPMN process models between BPMN~1.1 and BPMN~2.0 models. %\dnote{for consistency throughout the paper, I would use 'models' here instead of diagrams}

    \item Decision Model and Notation (DMN): DMN is a standardized notation for modeling business decisions, complementing BPMN~\cite{omgdmn}.

     \item Case Management Model and Notation (CMMN): CMMN is an attempt to supplement BPMN and DMN with a notation that focuses on agility and autonomy~\cite{omgcmmn}.

    \item Event-driven Process Chain (EPC): EPC~\cite{doi:https://doi.org/10.1002/0471741442.ch6} is a process modeling notation that enjoyed substantial popularity before the advent of BPMN.

    \item Unified Modeling Language (UML): UML is a modeling language used to describe software (and other) systems. It is subdivided into class and use case diagrams.
 
    \item Value Chain: A value chain is an informal notation for sketching high-level end-to-end processes and process frameworks.

    \item ArchiMate: ArchiMate is a notation for the integrated modeling of information technology and business perspectives on large organizations~\cite{10.1007/978-3-642-01862-6_30}.

    \item Organization Chart: Organization charts are tree-like models of organizational hierarchies.

    \item Fundamental Modeling Concepts (FMC) Block Diagram: FMC block diagrams support the modeling of software and IT system architectures.

    \item(Colored) Petri Net: Petri nets~\cite{petri1962kommunikation} are a popular mathematical modeling language for distributed systems and a crucial preliminary for many formal foundations of BPM. In \datasetacr, colored Petri nets~\cite{jensen1987coloured} are considered a separate notation.

    \item Journey Map: Journey maps model the customer's perspective on an organization's business processes.

    \item Yet Another Workflow Language (YAWL): YAWL is a language for modeling the control flow logic of workflows~\cite{VANDERAALST2005245}.

    \item jBPM: jBPM models allowed for the visualization of business process models that could be executed by the jBPM business process execution engine before the BPMN 2.0 XML serialization format existed. However, recent versions of jBPM rely on BPMN 2.0-based models.

    \item Process Documentation Template: Process documentation templates support the generation of comprehensive PDF-based process documentation reports. These templates are technically a model notation, although they may practically be considered a reporting tool instead.

    \item XForms: XForms is a (dated) standard for modeling form-based graphical user interfaces~\cite{xforms}. 

    \item Chen Notation: Chen notation diagrams~\cite{10.1145/320434.320440} allow for the creation of entity-relationship models.
\end{itemize}

\datasetacr\ is available at \url{https://zenodo.org/record/7012043}. Its license supports non-commercial use for research purposes, e.g., usage for the evaluation of academic research artifacts, such as algorithms and related software artifacts.

\section{Properties of \datasetacr}
%%%%%%%%%%%%%%%%%%%%%%%%%%%%%

%\datasetacr\ comprises 1,021,475 models of varying complexity and in different modeling notations and languages. This includes vendor-provided examples, which are automatically added to workspaces upon their creation \footnote{About 470,000 models in the dataset use the names of the example processes as process names. However, this can only be used as a rough estimate of the number of example processes in the dataset.}. \dnote{I'm not sure if we should move the paragraph about the total number of models to originis of SAP-SAM}
%When excluding the examples, 549975 models remain in the dataset \dnote{I would rather not mention this number here since we decided that based on the process name which is more of a rough estimation than a precise number}.
%When considering each saved \emph{version/revision} of a model as a separate object, this results in a total of~2,207,983 models. TK: we won't have revisions by the deadline, but may re-add this for the camera-ready

\datasetacr\ comprises models in different modeling notations and languages, as well as of varying complexity. In this section, we provide an overview of selected properties of \datasetacr. The source code that we used to examine the properties is available at \url{https://github.com/signavio/sap-sam}.

\begin{figure}[hbt]
%\vspace{-2em}
    \centering
    \includegraphics[width=\linewidth]{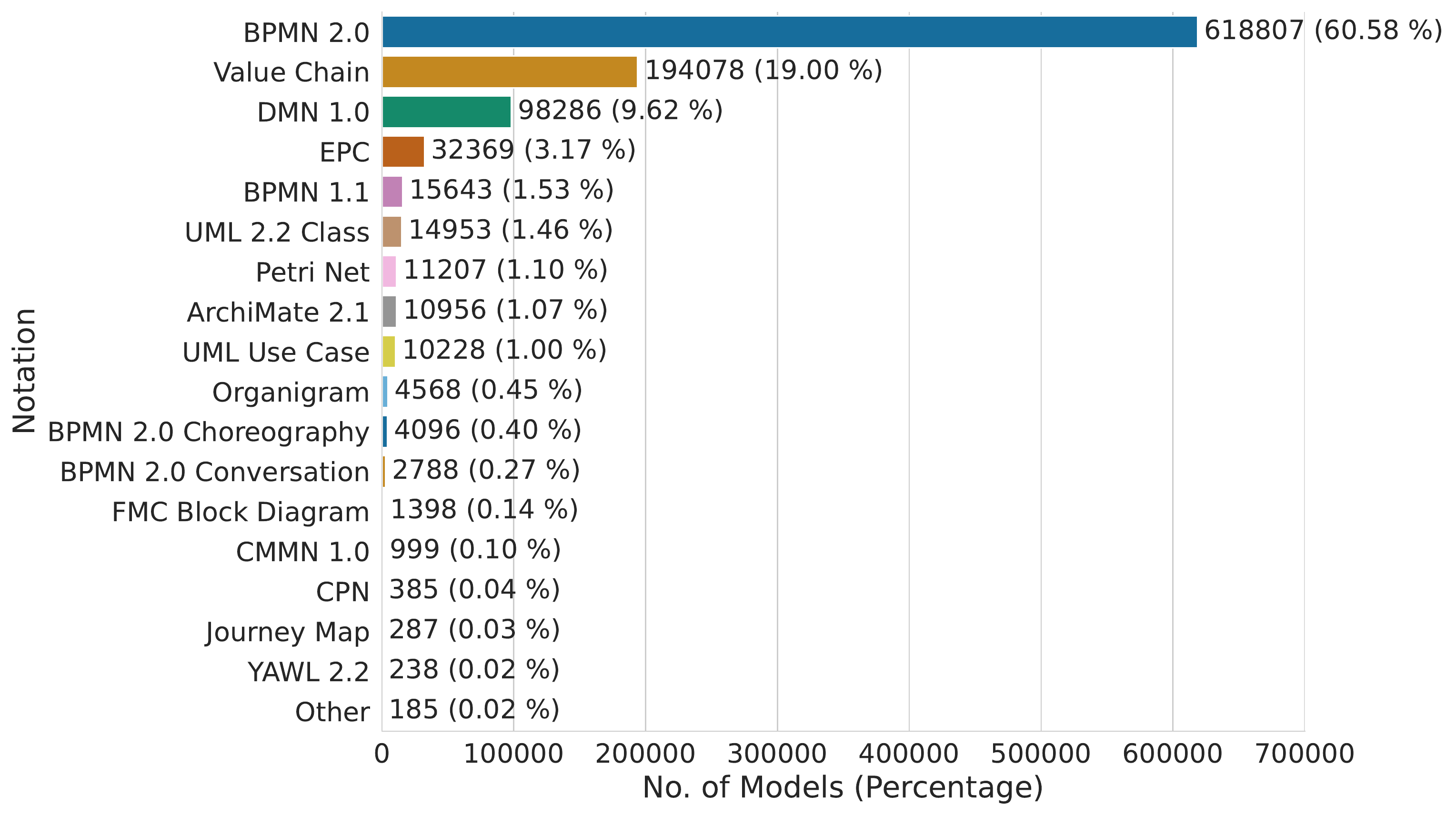}
    \caption{Usage of different modeling notations.}
    \label{fig:notations}
%\vspace{-2em}
\end{figure}

\mypar{Modeling notations}
\autoref{fig:notations} depicts the number of models in different notations in the dataset, as well as the according percentages (in brackets). We aggregate notations which are used for less than 100 models respectively into \textit{Other}: Process Documentation Template (86 models), jBPM 4 (76 models), XForms (20 models), and Chen Notation (3 models). The primarily used modeling notation is BPMN 2.0, which confirms that it is the de-facto standard for modeling business processes~\cite{chinosi2012bpmn}. Therefore, we will focus on BPMN 2.0 models as we examine further properties.

%The creation of models in the five most frequently used notations over time is visualized by \autoref{fig:notations-time}.

%\begin{figure}[!htbp]
%\vspace{-2em}
%    \centering
%    \includegraphics[width=\linewidth]{figures/notations_over_time.pdf}
%    \caption{Creation of models in different modeling notations over time. Note that the dataset only includes models created not later than September 2021.}
%    \label{fig:notations-time}
%\vspace{-2em}
%\end{figure}

%\begin{itemize}
%    \item statistics: percentages of different notations
%    \item if feasible: visualize the creation of models by notation over time 
%    \item argument that we concentrate on the predominant BPMN 2.0 models in the following
%    \item highlight journey models or other notations that Signavio wants to push (use of notations industry vs. academia)
%\end{itemize}

\begin{figure}[t]%htb
%\vspace{-1em}
    \centering
    \includegraphics[width=\linewidth]{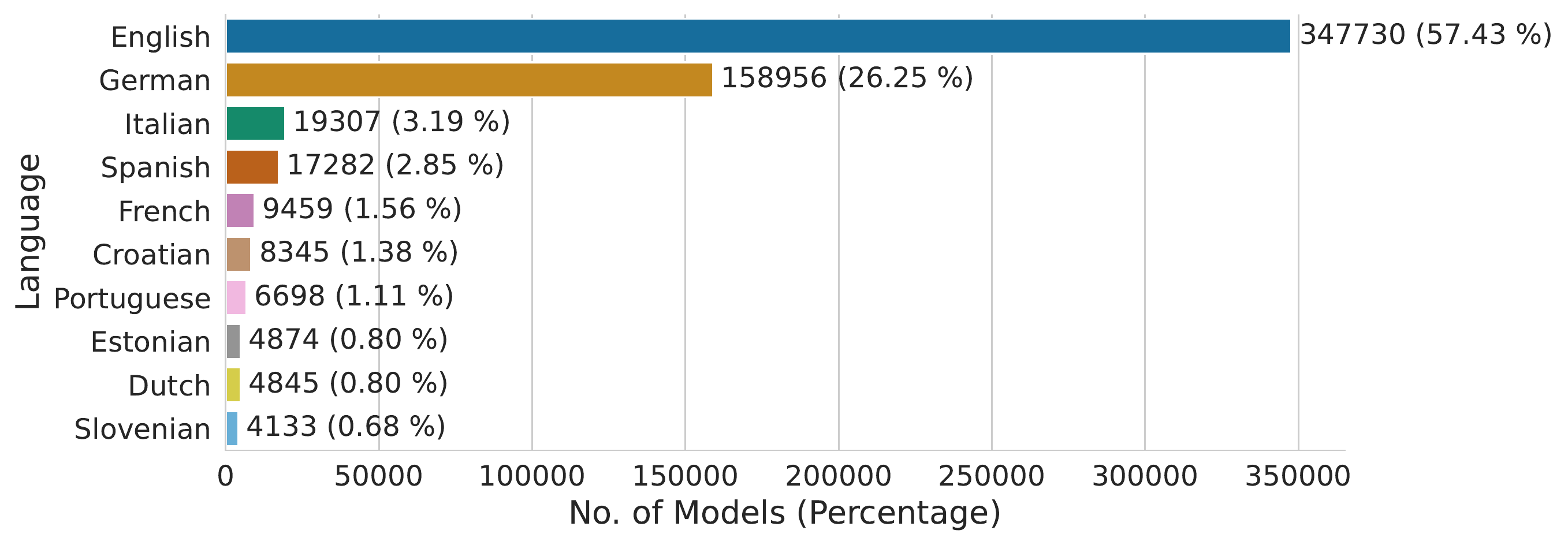}
    \caption{Usage of different languages for BPMN 2.0 models.}
    \label{fig:natural-langs}
%\vspace{-1em}
\end{figure}

\mypar{Languages}
Since \saiacr\ can be used by academic researchers, teachers and students all over the world, the models in \datasetacr\ are created using different languages. For example, \datasetacr\ includes BPMN~2.0 models in 41 different languages. \autoref{fig:natural-langs} shows the ten most frequently used languages for BPMN~2.0 models. Note that the vendor-provided example models, which are added to newly created workspaces, exist in English, German, and French. When a \saiacr\ workspace is created, the example models added to it are in German or French if the language configured upon creation is German or French, respectively; otherwise, the example models are in English. This contributes to the fact that more than half of the BPMN 2.0 models (57.43~\%) are in English.

\begin{figure}[t]%[!h]
%\vspace{-1em}
    \centering
    \includegraphics[width=.98\linewidth]{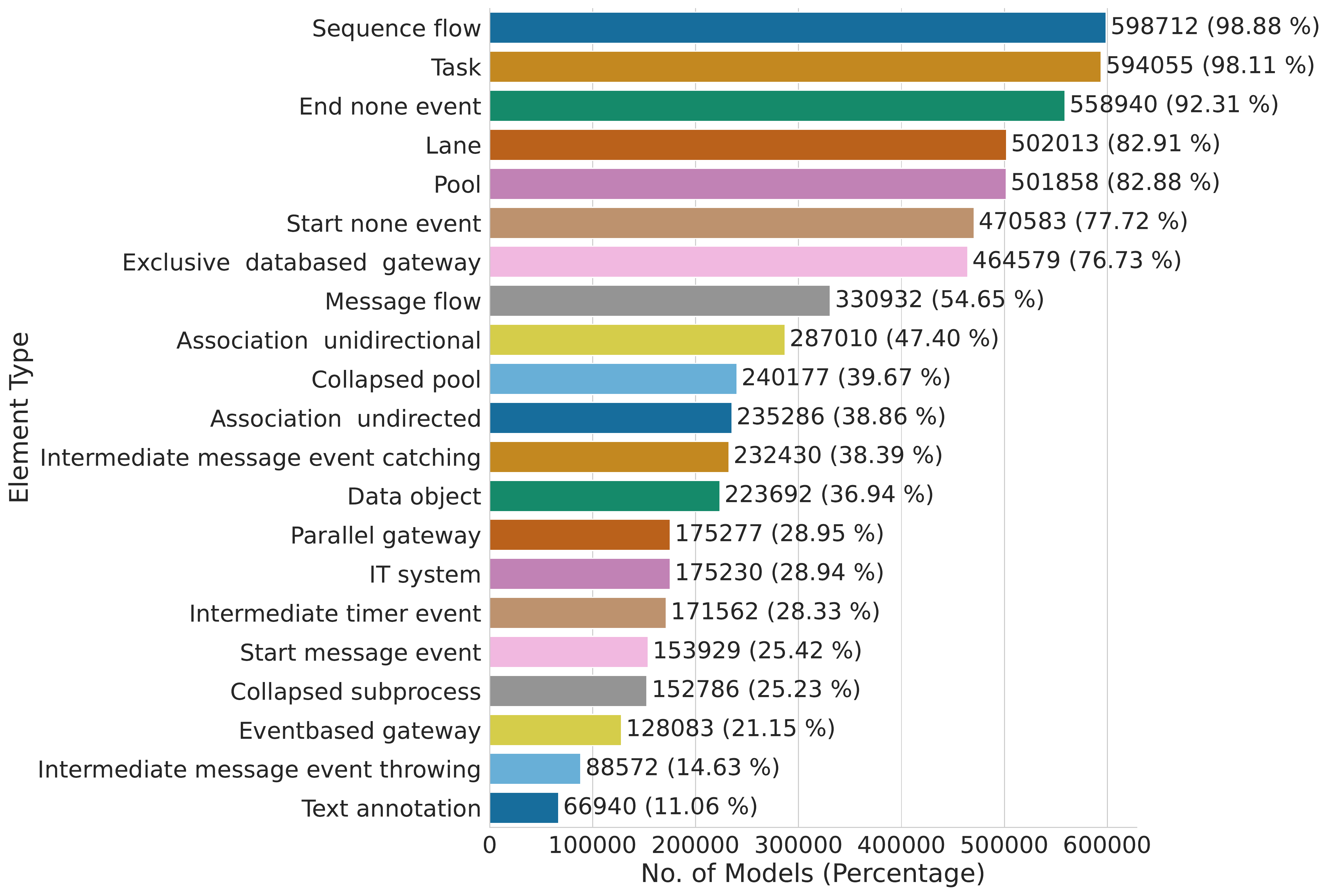}
    \caption{Occurrence frequency of different BPMN 2.0 element types.}
    %\Description[An uncompleted business process model]{An uncompleted business process model showing a version of the order-to-cash process from customer's perspective in a BPMN similar notation.}
    \label{fig:element-usage}
%\vspace{-1em}
\end{figure}

\setlength{\tabcolsep}{5pt}
\begin{table}[t]
\caption{Statistics of the number of elements per BPMN 2.0 model by type (grouped).}
\sisetup{table-format=2.1}
\centering
%\resizebox{0.8\columnwidth}{!}{
\begin{tabular}{l SSSSSSS[table-format=4.0]}
\toprule
 \multicolumn{1}{c}{Element type groups} & \multicolumn{1}{c}{Mean} & \multicolumn{1}{c}{Std} & \multicolumn{1}{c}{Min} & \multicolumn{1}{c}{25\%} & \multicolumn{1}{c}{50\%} & \multicolumn{1}{c}{75\%} & \multicolumn{1}{c}{Max} \\
 \midrule
Activities         &   8.6 &   8.4 &  0 &   4 &   7 &  10 &  1543 \\
Events             &   5.2 &   5.1 &  0 &   2 &   5 &   6 &   157 \\
Gateways           &   3.7 &   4.4 &  0 &   2 &   3 &   4 &   303 \\
Connecting Objects &  23.1 &  21.8 &  0 &  14 &  20 &  25 &  2066 \\
Swimlanes          &   3.8 &   2.6 &  0 &   3 &   4 &   5 &   227 \\
Data Elements      &   1.3 &   3.4 &  0 &   0 &   0 &   2 &   266 \\
Artifacts          &   0.9 &   4.0 &  0 &   0 &   0 &   1 &   529 \\
\bottomrule
\end{tabular}
%}

\label{tab:statistics-elements}
%\vspace{-1em}
\end{table}

\mypar{Elements}
\autoref{fig:element-usage} illustrates the occurrence frequency of different element types in the BPMN 2.0 models of \datasetacr.
It can be recognized that the element types are not equally distributed, which confirms the findings of prior research~\cite{muehlen2013much}. 
The number of models that contain at least one instance of a particular element type is much higher for some types, e.g., sequence flow (98.88~\%) or task (98.11~\%), than for others, e.g., collapsed subprocess (25.23~\%) or start message events (25.42~\%). Note that \autoref{fig:element-usage} only includes element types that are used in at least 10~\% of the BPMN~2.0 models. More than~30 element types are used by less than~1~\% of the models. On average, a BPMN 2.0 model in \datasetacr\ contains~11.3 different element types (median:~11) and 46.7 different elements, i.e., instances of element types (median:~40).

\autoref{tab:statistics-elements} shows the number of elements per model by type. For a compact representation, we aggregate similar element types by arranging them into groups. On average, connecting objects, which include associations and flows, make up the largest proportion of the elements in a model (mean:~23.1, median:~20).

% Not enough space:
%\subsection{Organisations}
%\begin{itemize}
%    \item \todoinline{see what Liz did so far}
%    \item distribution of no. of users per organisation
%    \item distribution of no. of created models per organisation
%    
%\end{itemize}

% Not enough time and space:
%\subsection{Syntax}
%\todoinline{eher für extension}
%\begin{itemize}
%    \item check if we can get interesting outputs from the syntax checker to write about that here (Simon)
%\end{itemize}

\begin{figure}[t]%[htb]
%\vspace{-2em}
    \centering
    \includegraphics[width=\linewidth]{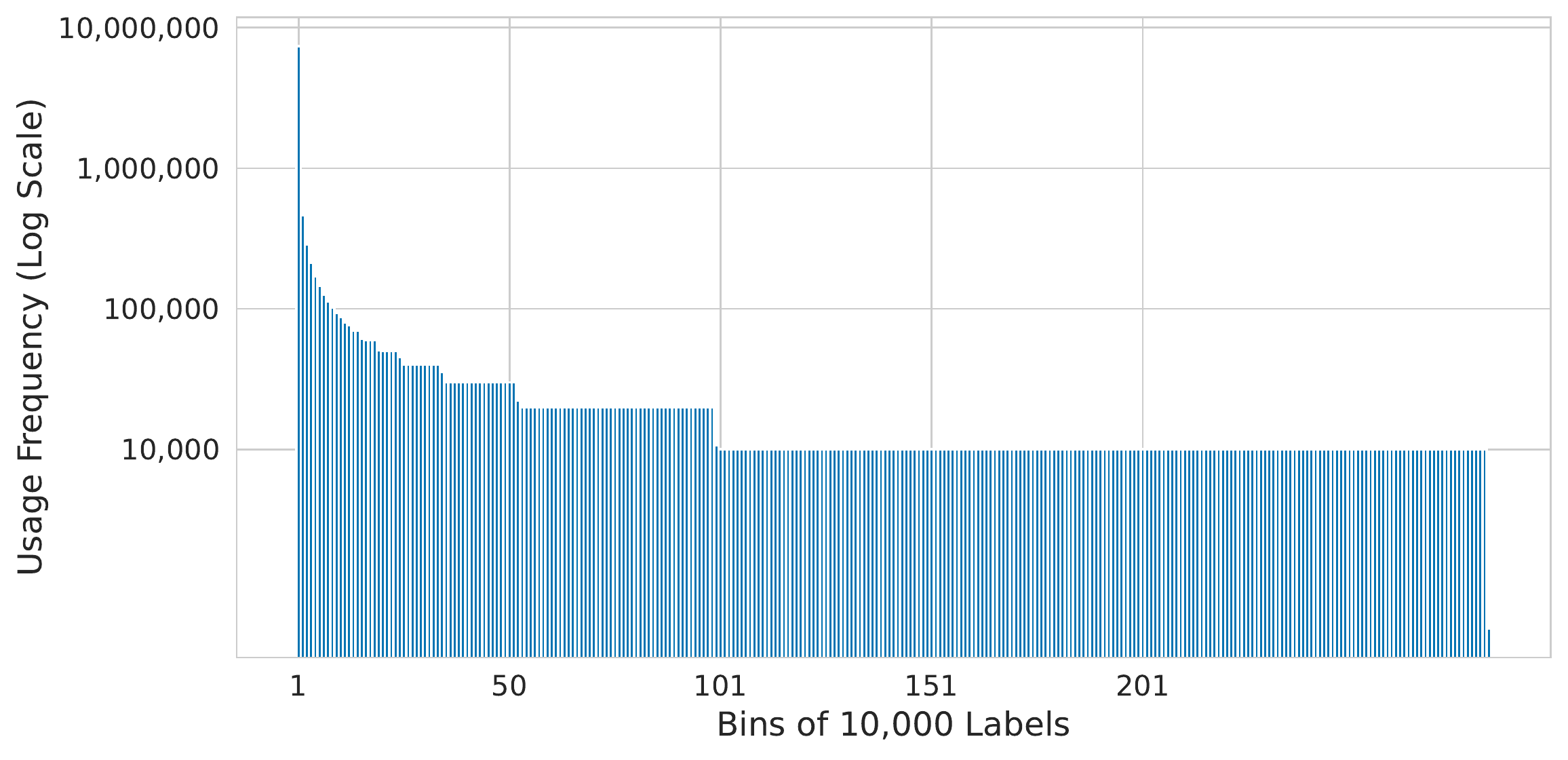}
     \caption{Distribution of the label usage frequency in BPMN 2.0 models. Each bar represents a bin of 10,000 distinct labels.}
    \label{fig:labels-usage}
%\vspace{-2em}
\end{figure}

\mypar{Labels} All elements of a BPMN 2.0 model can be labeled by the modeler, which results in a total of 2,820,531 distinct labels for the 28,293,762 elements of all BPMN 2.0 models in \datasetacr. \autoref{fig:labels-usage} depicts the distribution of label usage frequencies. We sorted the labels based on their absolute usage frequency in descending order and aggregated them in bins of size 10,000 to visualize the unevenness of the distribution. The first bin (leftmost bar in the chart) therefore contains the 10,000 most frequently used labels for the elements in the BPMN 2.0 models. Overall, 53.9~\% of all elements in the BPMN 2.0 models are labeled with these first 10,000 labels. On the other hand, the long-tail distribution indicates that many of the labels are used for only one element of all BPMN 2.0 models. More precisely, 1,829,891 (64.9~\%) of the labels are used only one time. The unevenness of the label usage distribution can again partly be explained by the vendor-provided examples in the dataset: The labels of the example processes appear very frequently in the dataset.

%%%%%%%%%%%%%%%%%%%%%%%%%%%%%%%%%%%%
\section{\datasetname\ Applications}
%%%%%%%%%%%%%%%%%%%%%%%%%%%%%%%%%%%%

As pointed out above, large process model collections like \datasetacr\ are a valuable and critical resource for BPM research. 
Process models from practice codify organizational knowledge about business processes and methodical knowledge about modeling practices. 
Both types of knowledge can be used by research, for example, for deriving recommendations for the design of future models.  
In addition, large process model collections are required for evaluating newly developed BPM algorithms and techniques regarding their applicability in practice.

To illustrate the potential value of \datasetacr\ for the BPM community, the following list describes some application scenarios that we consider to be particularly relevant.
It is neither prescriptive nor comprehensive; researchers can use \datasetacr\ for many other purposes.

\mypar{Knowledge Generation} 
    Process models depict business processes, codifying knowledge about the operations within organizations. 
    This knowledge can be extracted and generalized to a broader context. 
    Hence, \datasetacr\ can be considered as a knowledge base to generate new insights into the contents and the practice of organizational modeling. Example applications include:
    \begin{itemize}
        \item Reference model mining \cite{rehse2017graph}: Reference models provide a generic template for the design of new processes in a certain industry.
        They can be mined by merging commonalities between existing processes from different contexts into a new model that abstracts from their specific features. 
        By applying this technique to subsets of similar models from \datasetacr, we can mine new reference models for process landscapes or individual processes, including, e.g., the organizational perspective. 
        Similarly, we could identify, analyze, and compare different variants of the same process.
        \item Identifying modeling patterns \cite{fellmann2018business}: Process model patterns provide proven solutions to recurring problems in process modeling. They can help in streamlining the modeling process and standardizing the use of modeling concepts.
        A dataset like \datasetacr\, which contains process models from many different modelers, provides an empirical foundation both for finding new modeling patterns and for validating existing ones. 
        This also extends to process model antipatterns, i.e., patterns that should be avoided, as well as modeling guidelines and conventions.
    \end{itemize}
    
\mypar{Modeling Assistance} 
    The modeling knowledge that is codified in \datasetacr\ can also be used for automated assistance functions in modeling tools. Such assistance functions support modelers in creating or updating process models, accelerating and facilitating the modeling process. 
    However, many assistance functions are based on machine learning techniques and therefore require a large set of training data to generate useful results. 
    With its large amounts of contained modeling structures and labels,  \datasetacr\ offers a substantial training set, for example, for the following applications:
    \begin{itemize}
        \item Process model auto-completion \cite{sola2022exploiting}. By providing recommendations on possible next modeling steps, process model auto-completion can speed up modeling and facilitate consistency of the terms and modeling patterns that are used by an organization. Besides structural next element type recommendations, text label suggestions or even recommendations of entire process segments are possible. \datasetacr\ can be used to train machine learning models for these purposes. 
        \item Automated abstraction techniques \cite{wang2018business}: One important function of BPM is process model abstraction, i.e., the aggregation of model elements into less complex, higher-level structures to enable a better understanding of the overall process. 
        Such an aggregation entails the identification and assignment of higher-level categories to groups of process elements.  
        \datasetacr\ can provide the necessary training data for an NLP-based automated abstraction.
    \end{itemize}
    
\mypar{Evaluation} 
    Managing large repositories of process models is a key application of BPM \cite{dijkman2012managing}. Researchers have developed many different approaches to assist organizations with this task. 
    To make these approaches as productive as possible, they need to be tested on datasets that are comparable to those within organizations. 
    Since \datasetacr\ goes well beyond the size of related datasets, it can be used for large-scale evaluations of existing process management approaches on data from practice. 
    Examples for these approaches include process model querying \cite{Polyvyanyy2022}, process model matching \cite{antunes_2015_matching}, and process model similarity \cite{dijkman2011similarity}.

%%%%%%%%%%%%%%%%%%%%%%%%%%%%%%%%%%%%
\section{Limitations and Recommendations for Usage}
%%%%%%%%%%%%%%%%%%%%%%%%%%%%%%%%%%%%

As explained in the previous section, SAP-SAM can be used by the academic community to test and evaluate a plethora of tools and algorithms that address a wide range of process querying and business process analytics use cases.
However, in the context of any evaluation, the limitations of the dataset need to be taken into account. Considering the nature of \datasetacr\ as a model collection that has been generated by academic researchers, teachers, and students, the following limitations must be considered:
\begin{itemize} 
    \item Many models in \datasetacr\ exist multiple times, either as direct duplicates (copies) or as very similar versions. This includes vendor-provided example models or standard academic examples that are frequently used in academic teaching and research. The existence of these models can be used to evaluate variant identification and fuzzy matching approaches in process querying, but it negatively affects the diversity, i.e.,\ the breadth of the dataset.
    \item Many models may be of low technical quality, in particular the models that are created by ``process modeling beginners'', i.e.,\ early-stage students, for learning purposes. Although it can be interesting to analyze the mistakes or antipatterns in such models, flawed models can, for example, be problematic when using the dataset for generating modeling recommendations based on machine learning. Also, the mistakes that students make are most likely not representative of mistakes made by process modeling practitioners.
    \item Because many of the models have most likely been created for either teaching, learning, or demonstrating purposes, they presumably present a simplistic perspective on business processes. Even when assuming that all researchers, teachers, and students are skilled process modelers\footnote{Considering the previous point, that means even when focusing on the subset of the model collection that only entails models carefully created by skilled advanced students, teachers, and researchers.} and have a precise understanding of the underlying processes when modeling, the purpose of their models is typically fundamentally different from the purpose of industry process models. 
    Whereas academic models often emphasize technical precision and correctness,  industry models usually focus on a particular business goal, such as the facilitation of stakeholder alignment.
    %While the former purpose often emphasizes technical precision and correctness, the latter focuses on a particular business goal, such as the facilitation of stakeholder alignment \dnote{I think the 'former' purpose mentioned here is mentioned very implicitly only - could you please rewrite @Timmi? I would do it but I'm not sure if I'm getting this right}.
\end{itemize}

Let us note that this list may not be exhaustive; in particular, limitations that depend on a particular use case or evaluation scenario need to be identified by researchers who will use this dataset.
Still, it is also worth highlighting that the rather ``messy'' nature of the model collection reflects the reality of industrial data science challenges, in which a sufficiently large amount of high-quality data (or models) is typically not straightforwardly available \cite{houy2011business}; instead, substantial efforts need to be made to separate the wheat from the chaff, or to isolate use-cases in which the flaws in the data do not have an adverse effect on business value, or any other undesirable organizational or societal implications. 

\begin{figure}[t]%[!htbp]
%\vspace{-2em}
    \centering
    \includegraphics[width=0.5\linewidth]{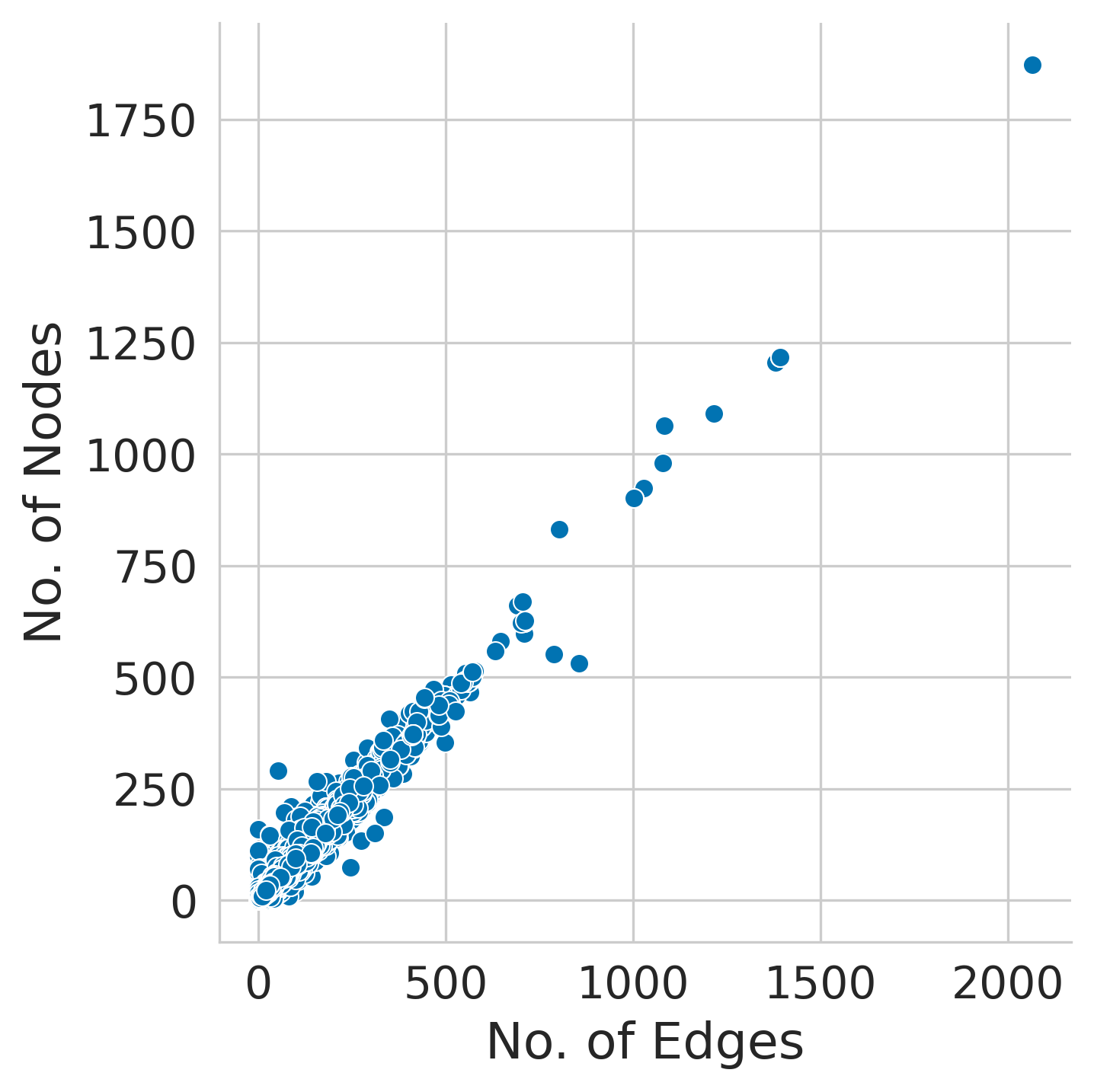}
    \caption{Correlation of the number of nodes and edges in BPMN 2.0 models.}
    %\Description[An uncompleted business process model]{An uncompleted business process model showing a version of the order-to-cash process from customer's perspective in a BPMN similar notation.}
    \label{fig:correlation}
%\vspace{-2em}
\end{figure}

When using \datasetacr\ for academic research purposes, it typically makes sense to filter it, i.e., to reduce it to a subset of models that satisfy desirable properties.
Here, we provide some recommendations to help with this step.
\begin{itemize}
    \item It typically makes sense to filter out the vendor-provided example models that are created by the \saiacr\ system upon workspace creation.
    %\item Depending on the use case, it can make sense to include all available model \emph{revisions}, \textit{i.e.}, versions of a particular models that reflect the different states in which a model was saved by the modeler(s). This can facilitate the study of change in process models. TK: Out-commended for now
    \item For many use cases, researchers may want to sort out process models that contain a very small or a very large number of elements. As can be expected for BPMN 2.0 models and is shown in \autoref{fig:correlation}, the number of nodes and the number of edges in a model are highly correlated. Hence, it is sufficient to filter according to the number of nodes. There is no need to additionally filter according to the number of edges. 
    \item Similarly, researchers may want to sort out process models where the element labels have an average length of less than, for example, three characters to ensure that only models with useful labels are included.
\end{itemize}
Let us again highlight that example code that demonstrates how the dataset can be queried, as well as the code for the analysis in this paper is available at \url{https://github.com/signavio/sap-sam}.

%%%%%%%%%%%%%%%%%%%%%%%%%%%%%%%%%%%
\section{Conclusion}
%%%%%%%%%%%%%%%%%%%%%%%%%%%%%%%%%%%

In this paper, we have presented the \datasetacr~dataset of process and other business models.
We are confident in our assumption that \datasetacr~is, by far, the largest publicly available collection of business process models.
Hence, it can---despite the limitation that it entails ``academic'' models created by researchers, teachers, and students and not by process management professionals---serve as an excellent basis for developing and evaluating tools and algorithms for process model querying and analysis.

In the future, SAP-SAM can potentially be augmented by including the following additional data objects:

\begin{itemize}
    \item Business objects/dictionary entries: In addition to models, \saiacr\ supports the creation of business objects, so-called \emph{dictionary entries}. These objects represent, for example, organizational roles, documents, or IT systems and can be linked to models to then be re-used across a process landscape that entails many models. Dictionary entries facilitate process landscape maintenance, as well as reporting.
    \item Standard-conform XML serializations: The models in the \datasetacr\ dataset are serialized using a non-standardized JSON format that \emph{i)} supports a generalization of modeling notations and \emph{ii)} is more convenient to use than XML-based serializations within the JavaScript-based front-ends of modern web applications. However, proprietary components exist that can---in the case of BPMN, DMN, and CMMN models---generate XML serializations which are compliant with the corresponding Object Management Group standards. Adding these XML serializations to the dataset can facilitate academic use, as many open-source and prototypical software tools support the open standards.
    \item PNG or SVG image representations: Similarly, to allow for a more straightforward visualization of models, PNG and SVG representations of the \datasetacr\ models can be generated and included.
\end{itemize}

%

\bibliographystyle{splncs04}
\bibliography{my.bib}
\end{document}